\documentclass[prb, onecolumn, nofootinbib, citeautoscript, 10pt, longbibliography, notitlepage]{revtex4-2}

\usepackage{graphicx}% Include figure files
\usepackage{dcolumn}% Align table columns on decimal point
\usepackage{bm}% bold math
\usepackage{amsmath}
\usepackage{tabularx,graphicx}
\usepackage{epstopdf}
\usepackage{latexsym}
\usepackage{amssymb}
\usepackage{color, colortbl}
\usepackage{psfrag}
\usepackage{bbm}
\usepackage{titlesec}
\usepackage{dsfont}
\usepackage{feynmp}
\usepackage{slashed}
\usepackage{multirow}
\usepackage{physics}
\usepackage[tight]{subfigure}

\usepackage[papersize={8.5in,11in}]{geometry}

\usepackage{color}
\definecolor{darkblue}{rgb}{0.,0.,0.4}
\definecolor{darkred}{rgb}{0.5,0.,0.}
\definecolor{BlueViolet}{RGB}{138,43,226}
\definecolor{SkyBlue}{RGB}{30,144,255}
\definecolor{DarkGreen}{RGB}{0,100,0}
\usepackage[pdftex,colorlinks=true,linkcolor=darkblue,citecolor=blue,urlcolor=darkred]{hyperref}

\def \nn{\nonumber \\}
\begin{document}

\title{Transmission through rectangular potentials in semimetals featuring quadratic dispersion}

\author{Ipsita Mandal}
\email{ipsita.mandal@snu.edu.in}

\affiliation{Department of Physics, Shiv Nadar Institution of Eminence (SNIoE), Gautam Buddha Nagar, Uttar Pradesh 201314, India}

\begin{abstract}
We revisit the problem of transmission of quasiparticles through a rectangular potential barrier, for semimetals featuring quadratic-in-momentum band-crossings at a nodal point. Although this was considered in Annals of Physics 419 (2020) 168235, the solutions corresponding to evanescent waves were missed, leading to a partial fulfillment of the boundary conditions required to determine the piecewise-continuous wavefunctions. In this paper, our aim is to correct those shortcomings, recompute the transmission coefficients, and show the resulting behaviour of the conductivity and the Fano factor for some representative parameter values.
\end{abstract}
\maketitle

\tableofcontents
%==========================================================================

\section{Introduction}
%%%%%%%%%%%%%%%%%%%%%%%%%%%%%%%%%%%%%

\begin{figure}[t]
{\includegraphics[width = 0.65 \textwidth]{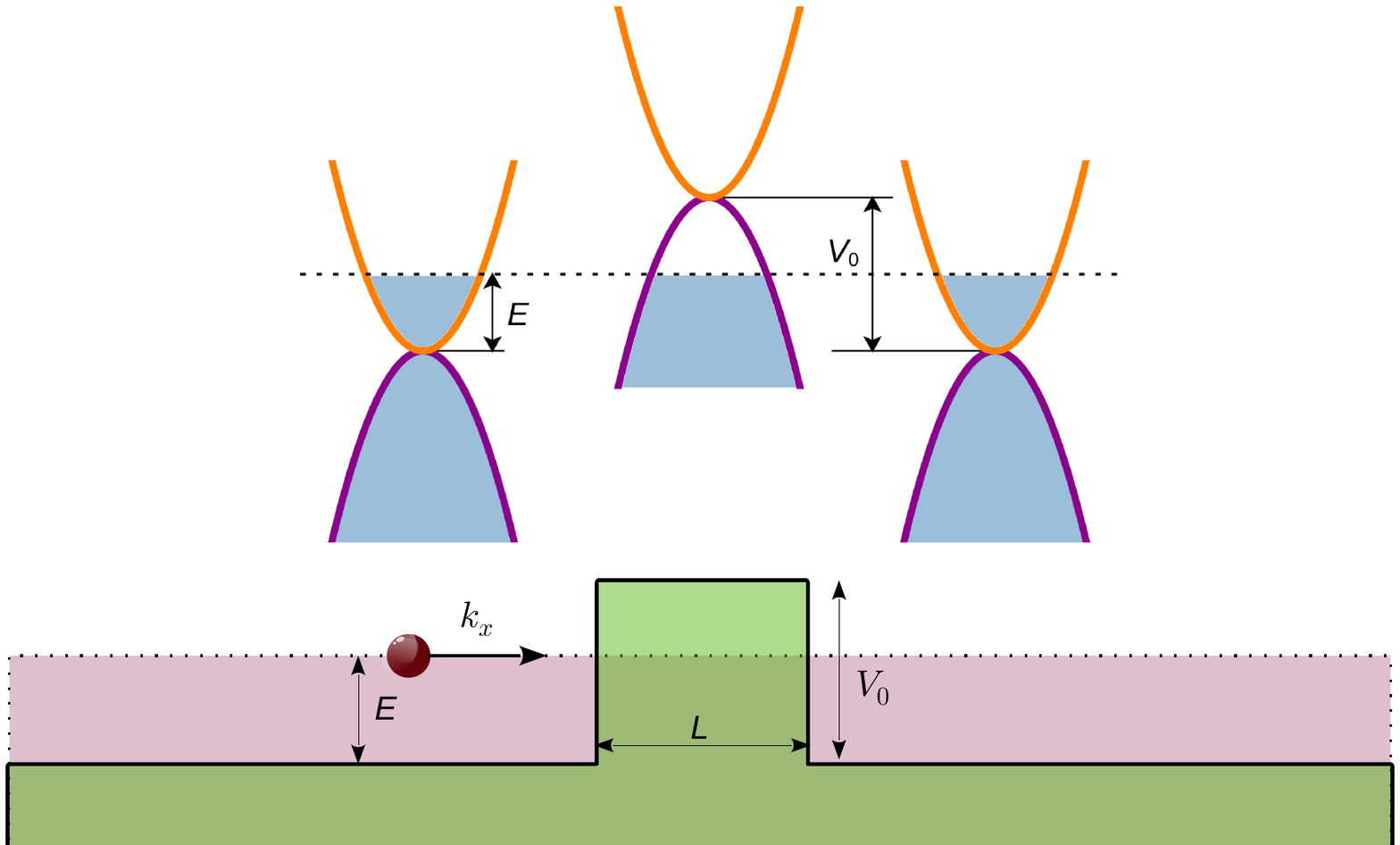}}
\caption{Schematics of transmission through a potential barrier in a system hosting a QBCP. The upper panel shows the schematic diagrams of the spectra of quasiparticles about a QBCP, when a potential barrier is encountered while moving along the $x$-direction. The lower panel represents the schematic diagram of the transport across the potential barrier. The Fermi level (indicated by dotted lines) lies in the conduction band outside the barrier, and in the valence band inside it. The blue-coloured areas of the bands indicate occupied states.}
\label{figbands}
\end{figure}

Nodal-point semimetals are materials which feature twofold or multifold band-crossing points in the Brillouin zone (BZ), where two or more bands cross. This implies that, when the chemical potential cuts the nodal point, the Fermi surface shrinks to a Fermi node. The poster-child examples of such materials comprise (1) graphene \cite{graphene_review}, for the case of two-dimensional (2d) systems; and (2) Dirac and Weyl semimetals \cite{armitage_review}, for the case of three-dimensional
(3d) systems. For each of these cases, the bands crossing at a node show linear-in-momentum dependence. However, there exist 2d and 3d semimetals where the bands crossing at a node show quadratic-in-momentum dependence \cite{kai-sun, tsai, ips-seb, Abrikosov, LABIrridate, MoonXuKimBalents, rahul-sid, ipsita-rahul, ips_tunnel_qbcp, armitage, ips-hermann1, ips-hermann2, ips-hermann3, ips_qbt_plasmons, ips-jing, ips-sandip}, which are sometimes dubbed as quadratic-band-crossing points (QBCPs). While 2d QBCPs are realised in checkerboard \cite{kai-sun} (at half-filling), Kagome \cite{kai-sun} (at one-third-filling), and Lieb \cite{tsai} lattices, various pyrochlore iridates, $\text{A}_2\text{Ir}_2\text{O}_7$ (where A stands for a lanthanide element~\cite{pyro1,pyro2}), host 3d QBCPs. Such bandstructures have also been realised in 3d gapless semiconductors in the presence of a sufficiently strong spin-orbit coupling \cite{Beneslavski}, such that the resulting model is of relevance for materials like gray tin ($\alpha$-Sn) and mercury telluride (HgTe). These systems are also known as ``Luttinger semimetals'' due to the fact that the low-energy quasipartcles (in the vicinity of a node) are captured by the Luttinger Hamiltonian of inverted band-gap semiconductors \cite{Abrikosov, luttinger, murakami, igor16}.
In this paper, we will compute the transmission and reflection coefficients of the quasiparticles in the vicinity of 2d and 3d QBCPs, when moving across a rectangular potential, which is chosen to lie in a path along the $x$-axis or $z$-axis. The set-up is represented schematically in Fig.~\ref{figbands}. Although this was considered in Ref.~\cite{ips_tunnel_qbcp}, the solutions corresponding to evanescent waves were missed (see Refs.~\cite{deng2020, ips-aritra, banerjee, ips-abs-semid} for analogous situations in other kinds of semimetals), leading to a partial fulfillment of the boundary conditions required to determine the piecewise-continuous wavefunctions. With the correct solutions in place, we will determine the transport-characteristics reflected in the reflection and transmission coefficients.
After determining the transmission amplitudes ($ T_{\mathbf n} $'s), we will compute the conductance ($G$), given by the Landauer formula of \cite{landauer, buttiker, blanter-buttiker}
\begin{align}
\label{eqland}
G = \frac{e^2} { 2\, \pi\, \hbar} 
\sum_{\mathbf n}
 T_{\mathbf n} \,.
\end{align}
%%%%%%%%%%%%%%%
Here, $e$ is the charge of one electron and $\mathbf n$ labels the transverse momentum modes in the the strip of the material (in the experimental set-up). When an external potential difference, $\Phi $, is applied across a circuit, shot noise is the physical quantity which provides a measure of the fluctuations of the electric current-density away from its average value. At zero temperature, while the actual shot noise is defined by 
$ \mathcal S =  \frac{e^3\, |\Phi| } {  \pi\, \hbar} 
\sum_{\mathbf n}
 T_{\mathbf n} \left( 1 - T_{\mathbf n} \right ) ,$
the Poisson noise is given by the expression
% blanter-buttiker review pg-27
$
\mathcal S_P =  \frac{e^3\, |\Phi| } {  \pi\, \hbar} 
\sum_{\mathbf n} T_{\mathbf n} $ \cite{blanter-buttiker}.
The Poisson noise is the value of the noise that would be measured if the system produced noise due to quasiparticles carrying a single set of the relevant quantum numbers. A convenient measure of the sub-Poissonian shot noise is the Fano factor ($F$), which is the ratio defined as \cite{blanter-buttiker}
\begin{align}
\label{eqfano}
F = \frac{\mathcal S}{ \mathcal S_P } =
\frac{\sum_{\mathbf n}
 T_{\mathbf n} \left( 1 - T_{\mathbf n} \right )}
 { \sum_{\mathbf {\tilde n}} T_{\mathbf{\tilde  n}} } \,.
\end{align}

The paper is organized as follows. Sec.~\ref{sec2dmodel} deals with 2d QBCPs and Sec.~\ref{sec3dmodel} investigates 3d QBCPs. Therein, the nature of conductivity and Fano factors are compared with those of other types of bandstructures. Finally, we conclude with a summary and outlook in Sec.~\ref{secsum}. In all our expressions, we will be using the natural units, which means that the reduced Planck's constant ($\hbar $), the speed of light ($c$), and the magnitude of a single electronic charge ($e$) are each set to unity. Throughout the paper, we will ignore the spin-degeneracy of the electronic quasiparticles.

%%%%%%%%%%%%%%%%%%%%%%%%%%%%%%%%%%%%%%%%%%%%%%%%%%%
%%%%%%%%%%%%%%%%%%%%%%%

%%%%%%%%%%%%%%%%%%%%%%%%%%%%%%%%%%%%%%%%%%%%%%%%%%
\begin{figure}[t]
\includegraphics[width = 0.75 \textwidth]{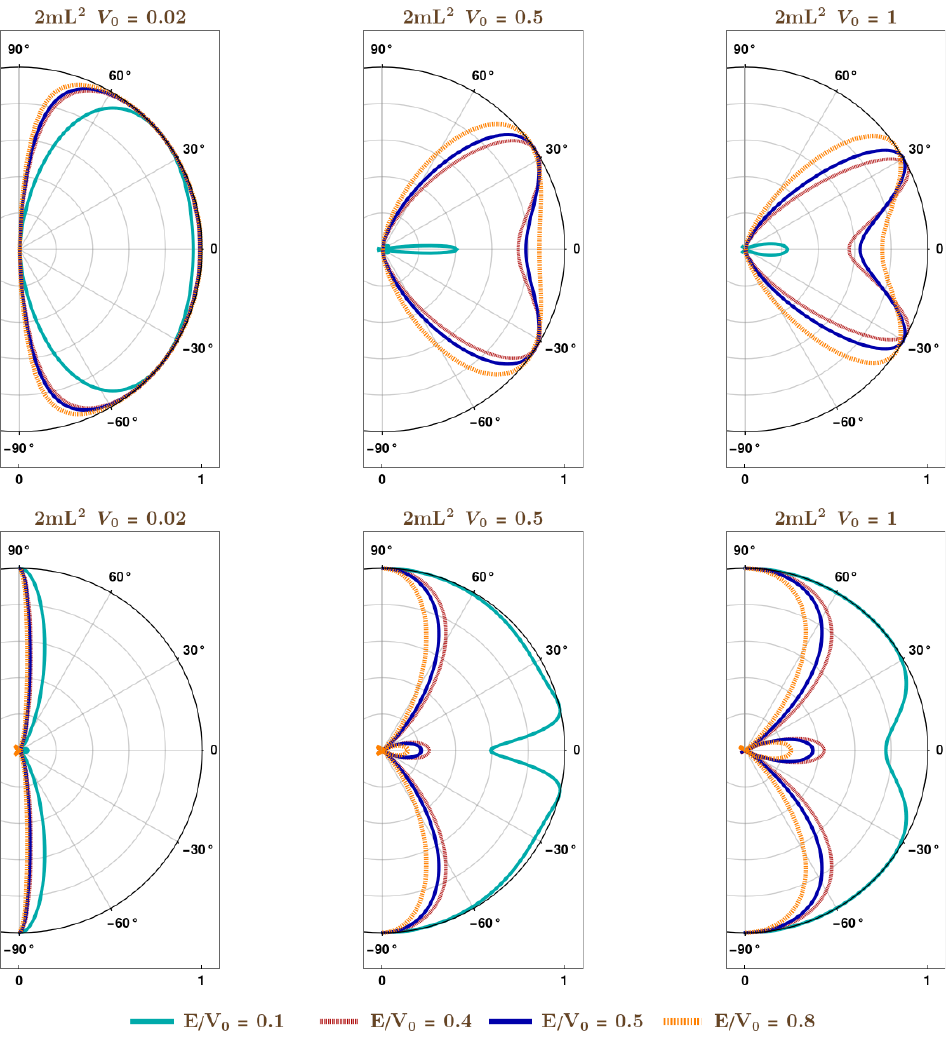}
\caption{2d QBCP: The polar plots show the transmission coefficient, $ T (E, V_0,\phi) \big \vert_{E < V_0}$ (upper panel), and the reflection coefficient $ R (E, V_0,\phi) \big \vert_{E \leq V_0}$ (lower panel), as functions of the incident angle $\phi$, for some representative values of $2\, m\, L^2\,V_0$. The values of $E$ are shown in the plot-legends.
\label{fig2dless}}
\end{figure}
%%%%%%%%%%%%%%%%%%%%%%%%

%%%%%%%%%%%%%%%%%%%%%%%%%%%%%%%%%%%%%%%%%%%%%%%%%%
\begin{figure}[t]
\includegraphics[width = 0.75 \textwidth]{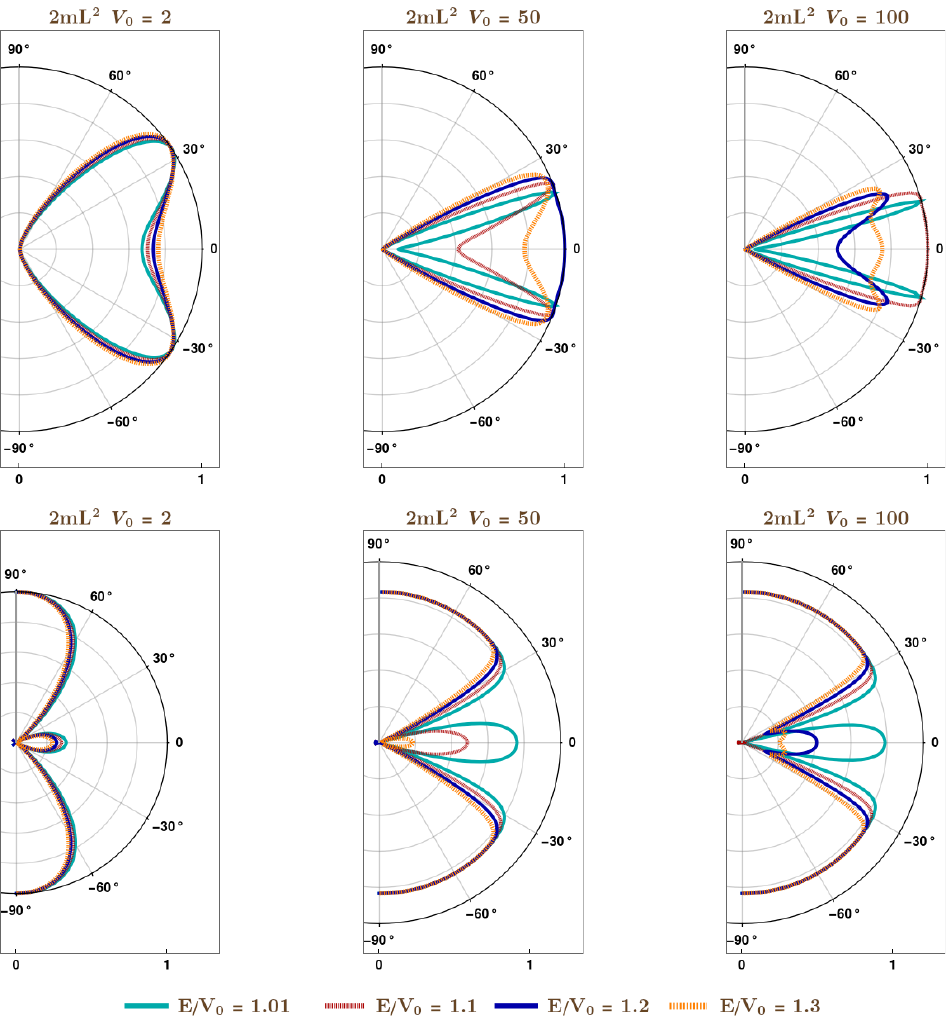}
\caption{2d QBCP: The polar plots show the transmission coefficient, $ T(E, V_0,\phi) \big \vert_{E > V_0}$ (upper panel), and the reflection coefficient $ T (E, V_0,\phi) \big \vert_{E > V_0}$ (lower panel), as functions of the incident angle $\phi$, for some representative values of $2\, m\, L^2\,V_0$. The values of $E$ are shown in the plot-legends.
\label{fig2dmore}}
\end{figure}
%%%%%%%%%%%%%%%%%%%%%%%%

%%%%%%%%%%%%%%%%%%%%%%%%%%%%%%%%%%%%%%%%%%%%%%%%%%
\begin{figure}[t]
\includegraphics[width = 0.75 \textwidth]{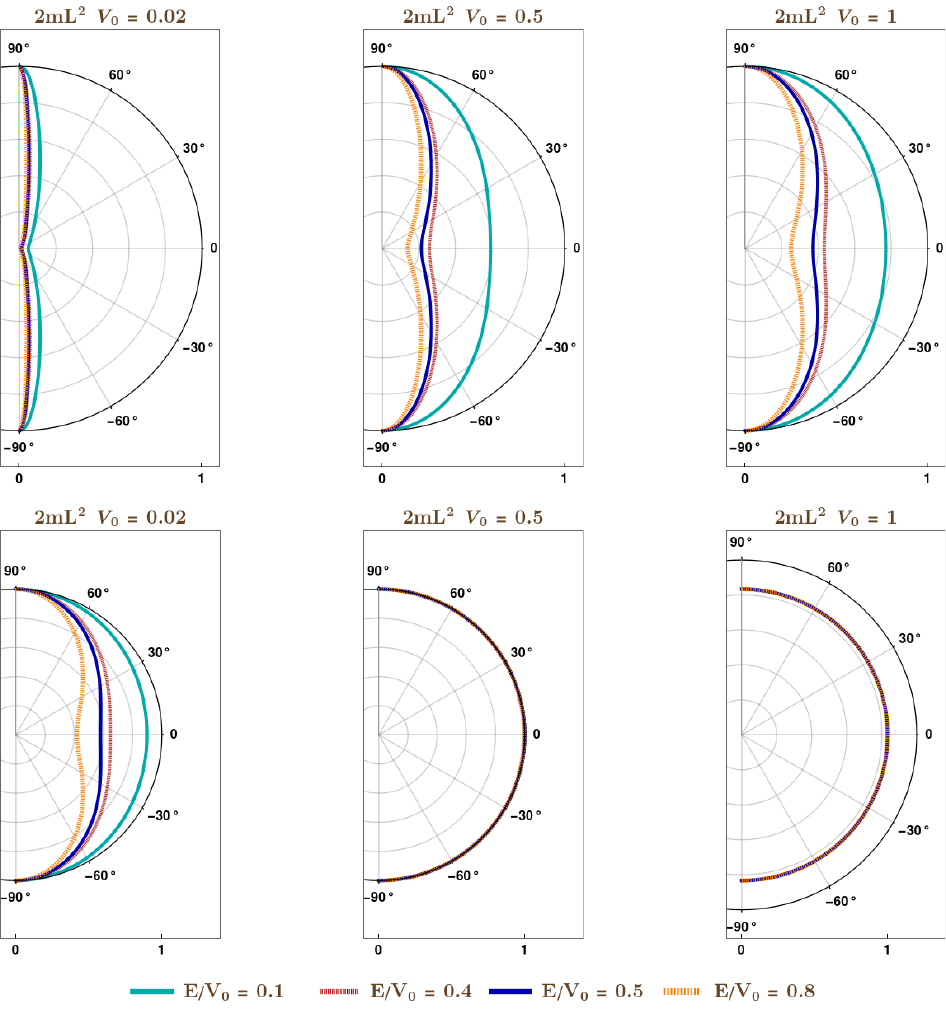}
\caption{2d electron-gas: The polar plots show the transmission coefficient, $ T (E, V_0,\phi) \big \vert_{E < V_0}$ (upper panel), and the reflection coefficient $ R (E, V_0,\phi) \big \vert_{E \leq V_0}$ (lower panel), as functions of the incident angle $\phi$, for some representative values of $2\, m\, L^2\,V_0$. The values of $E$ are shown in the plot-legends.
\label{fig2delecless}}
\end{figure}
%%%%%%%%%%%%%%%%%%%%%%%%

%%%%%%%%%%%%%%%%%%%%%%%%%%%%%%%%%%%%%%%%%%%%%%%%%%
\begin{figure}[t]
\includegraphics[width = 0.75 \textwidth]{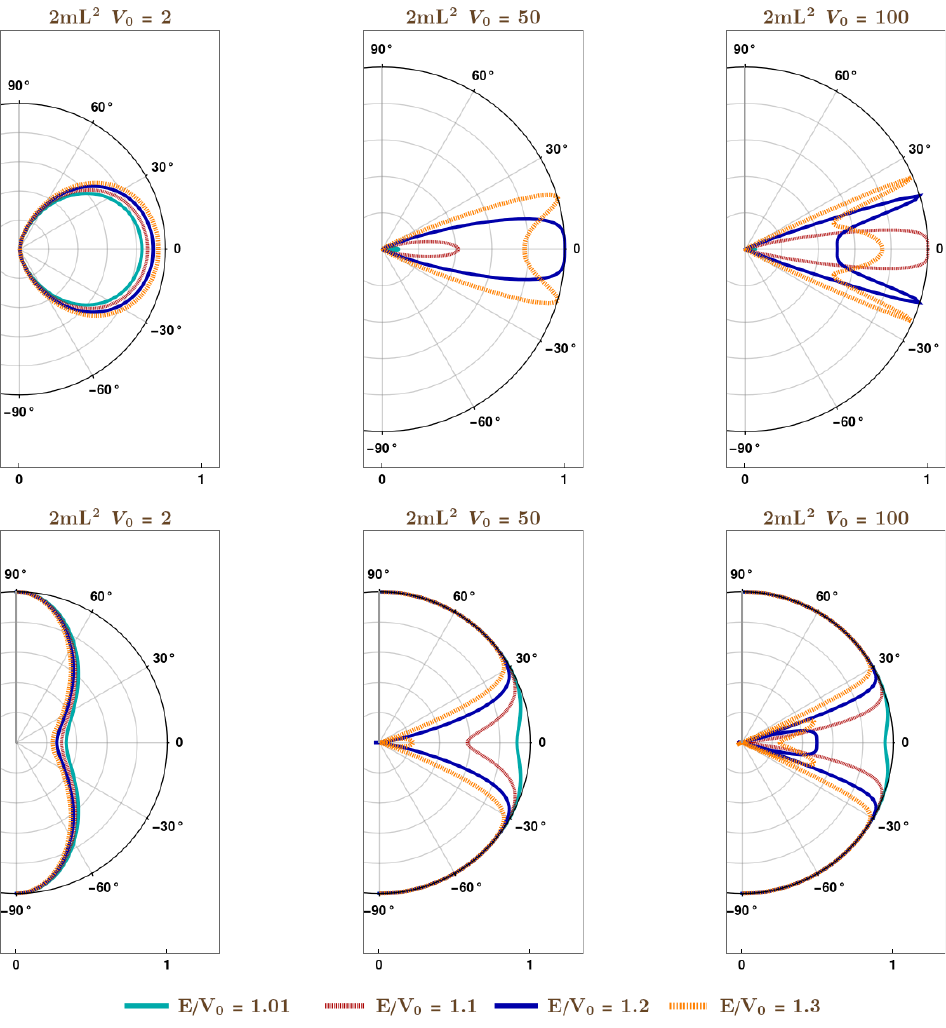}
\caption{2d electron-gas: The polar plots show the transmission coefficient, $ T(E, V_0,\phi) \big \vert_{E > V_0}$ (upper panel), and the reflection coefficient $ T (E, V_0,\phi) \big \vert_{E > V_0}$ (lower panel), as functions of the incident angle $\phi$, for some representative values of $2\, m\, L^2\,V_0$. The values of $E$ are shown in the plot-legends.
\label{fig2delecmore}}
\end{figure}
%%%%%%%%%%%%%%%%%%%%%%%%

%%%%%%%%%%%%%%%%%%%%%%%%%%%%%%%%%%%%%%
\section{2d QBCP}
\label{sec2dmodel}

A QBCP in a 2d system harbours a twofold band-crossing, represented by the low-energy effective Hamiltonian \cite{kai-sun},
\begin{align}
\mathcal{H}_{2d}^{kin}(k_x,k_y)&=
 \frac{1 }{2\,m}
 \Big [ \, 2\, k_x\,k_y\, \sigma_x +  \left( k_y^2-k_x^2\right) \sigma_z  \,\Big ],
\end{align} 
in the momentum space. The energy eigenvalues are captured by $\pm \,\varepsilon_{2d}(k_x,k_y)$, where
\begin{align}
\varepsilon_{2d}(k_x,k_y) =  \frac{   k_x^2 + k_y^2} {2\,m} \,,
\end{align}
where the ``$+$" and ``$-$" signs, in the usual conventions, refer to the positive-energy (or conduction) and negative-energy (or valence) bands, respectively.
The corresponding orthonormal eigenvectors are given by
\begin{align}
\Psi_+^T =\frac{1}{\sqrt{k_x^2+k_y^2}} \left( k_y  \quad  
k_x \right ) \text{ and }
  \Psi_-^T =\frac{1}{\sqrt{k_x^2+k_y^2}}  \left(  - \, k_x  \quad  
   k_y  \right ) ,
\end{align}
respectively.

The 2d system is modulated by a rectangular electric potential barrier of
height $V_0 $ and width $L$ [cf. Fig.~\ref{figbands}], giving rise to an $x$-dependent potential-energy function of the form
\begin{align}
V ( x ) = \begin{cases} V_0 &  \text{ for } 0 \leq x \leq L \\
0 & \text{ otherwise} 
\end{cases} .
\label{eqpot}
\end{align}
Hence, we need to consider the total Hamiltonian,
\begin{align}
\mathcal{H}_{2d}^{tot} &=
 \mathcal{H}_{2d}^{kin}(- \, i \,\partial_x \,,  \, -\, i\,\partial_y)
 + V(x) \,,
\end{align} 
in the position space. As mentioned earlier, here, the transport direction is chosen along the $x$-axis. We fix the Fermi-energy level at a value of $E >0$ in the region outside the potential barrier, which can be adjusted in an experimental set-up by either chemical doping or an external gate-voltage.

%%%%%%%%%%%%%%%%%%%%%%%%%%%
\subsection{Formalism}

For a material of a sufficiently large transverse dimension, $W$, the boundary conditions should be irrelevant for the bulk response, and we use this freedom to simplify the calculations. Here, on a physical wavefunction, $\Psi^{\rm{tot}} (x,y) $, we impose the periodic boundary condition of
$ \Psi^{\rm{tot}}(x,W) = \Psi^{\rm{tot}}(x,0) \,.$
The transverse momentum, $k_y$, is conserved, and it is quantized due to the periodicity in the transverse width $W$. Hence, it takes the form of $  k_y =\frac { 2\,\pi\,n} {W} \equiv q_n $,
where $n \in \mathbb{Z}$ and $ |k_y| \leq \sqrt{2\, m\, E } $. Then the full wavefunction is given by
\begin{align}
 \Psi^{\rm{tot}}(x,y,q_n) =
\text{const.}
\times \Psi_{ n}(x)\,  e^{  i\,q_n y }\,.
\end{align}
%%%%%%%%%%%%%%%%%%%%%%%%
For any mode of a given transverse-momentum component $ q_n $ (with $ | q_n | \leq 2\, m \,E   $), we can determine the $x$-component of the wavevectors of the incoming, reflected, and transmitted waves, by solving $
4 \, m^2 \, E^2 = \left( k_x^2 + q_n^2 \right)^2 \Rightarrow k_x^2
 = \pm \, 2 \, m\, E - q_n^2 \,.$
%%%%%%%%%%%%%%%%%%%%%%%
In fact, in the regions $x<0$ and $x>L$, this relation leads to the following four solutions: 
\begin{align}
\label{eqksol}
k_x = \pm \,\sqrt{2 \, m\, E - q_n^2} \text{ and }
k_x = \pm \, i\, \sqrt{2 \, m\, E + q_n^2} \,.
\end{align}
%%%%%%%%%%%%%%%%%%%%%%%
Consequently, in addition to the propagating plane-wave solutions, there are also evanescent waves present~\cite{banerjee,  deng2020, ips-aritra, ips-abs-semid}, characterized by complex values.\footnote{In general, if the Fermi energy cuts the bands at an energy $ \mu $, for propagation along the $x$-direction, the corresponding ``right-moving'' plane waves will have the factor $e^{i\, \text{sgn}( \mu ) \,k_x \, x}$. This just implies that, if the propagating quasiparticles are occupying the upper (lower) band, they have a positive (negative) group-velocity.} For $x < 0 $, we need to consider the solutions 
$ k_x = - \,\sqrt{2 \, m\, E - q_n^2} \text{ and }
k_x = - \, i\, \sqrt{2 \, m\, E + q_n^2} $ while determining the overall reflected wave, which clearly comprises evanescent parts. For $x >L $, we need to consider the solutions 
$ k_x = \sqrt{2 \, m\, E - q_n^2} \text{ and }
k_x =  i\, \sqrt{2 \, m\, E + q_n^2} $ while determining the overall transmitted wave, which again comprises evanescent parts. Inside the potential-barrier region, the $x$-components of the momentum are given by $ k_x =  \pm \, \sqrt{ 2\,m\,|E-V_0| -q_n^2} $ and $ k_x =  \pm \, i\, \sqrt{ 2\,m\,|E-V_0| + q_n^2} \,$. This situation resembles the case of bilayer graphene \cite{geim}.

We follow the usual procedure of matching of matching the wavefunctions across the piecewise-continuous regions (see, for example, Refs.~\cite{salehi,beenakker}) to compute the reflection and transmission coefficients. Here, we consider the transport of positive energy states (i.e., $\Psi_+$) corresponding to electron-like particles. The transport of hole-like excitations (i.e., $\Psi_-$) will be similar. Hence, the Fermi level outside the potential barrier is adjusted to the value $E $ (with $E>0$).
A scattering state $\sim \Psi_n (x) \, e^{i \,q_n y}$, labelled by the index $n$, is
constructed in a piecewise fashion as
%%%%%%%%%%%%%%%%%%%
\begin{align}
& \Psi_n (x)  =  \begin{cases} \phi_L (x)  & \text{ for } x<0   \\
 \phi_M  (x) & \text{ for } 0 \leq x \leq L  \\
\phi_R (x)  &  \text{ for } x > L 
\end{cases} ,
\end{align}
%%%%%%%%%%%%%%%%%%%%%%%%%%%%%%%%%%%%%%%%%%%%5
where
\begin{align}
 & \phi_L (x)   =  \frac{ 
 \Psi_+ (  k_{\rm in},  q_n) \,  e^{ i\, k_{\rm in} \, x }
+  r_n \, \Psi_+ (  - \, k_{\rm in},  q_{n}) \,   e^{- i\, k_{\rm in} \, x }
} 
{\sqrt{ \mathcal{V} (k_{\rm in},  q_n)}}
+
\tilde r_n \, \Psi_- (  - \,i \, \kappa ,  q_n) \,   e^{ - \kappa \,|x| } \, ,
 %%%%%%%%%
\nn  & \phi_M (x)   \nn & 
 =
  \Big[  \alpha_n\,\Psi_+ (  k_{\rm mid},  q_n ) \,
 e^{ i\,k_{\rm mid}\,  x } 
 + \beta_n \,\Psi_+ ( - \, k_{\rm mid},  q_n ) \,
 e^{- i\,k_{\rm mid}  \, x } 
+ 
\tilde \alpha_n \,  
\Psi_- (  i\, \kappa_{\rm mid},  q_n ) \, e^{-\, \kappa_{\rm mid}\,  x } 
+ 
\tilde \beta_n \,  
\Psi_- (  -\,i\,  \kappa_{\rm mid},  q_n ) \, e^{  \kappa_{\rm mid}\,  x } 
 \Big]  
 \nn & \qquad \times  \Theta \left( E-V_0  \right)
%%%%%%%%%%%%%%%
\nn &  \quad  + 
\Big[ \alpha_n\,\Psi_- (  k_{\rm mid},  q_n ) \,
 e^{ i\,k_{\rm mid}\,  x } 
 + \beta_n \,\Psi_- ( -\, k_{\rm mid},  q_{n}) \,
 e^{- i\,k_{\rm mid}  \, x }   
+ 
\tilde \alpha_n \,  
\Psi_+ (  i\, \kappa_{\rm mid},  q_n ) \, e^{ - \kappa_{\rm mid}\,  x } 
+ 
\tilde \beta_n \,  
\Psi_+ (  -\,i\,  \kappa_{\rm mid},  q_n ) \, 
e^{   \kappa_{\rm mid}\,  x }  \Big]   
\nn & \qquad \times \Theta \left( V_0 - E  \right) , \nn 
 %%%%%%%%%%%%%%%%%%%%%%%%%%%%%%%%%%%%%%%%%%%%%%%%%%%%%%%%%%%55
& \phi_R (x)  = 
  t_n\, \Psi_+ ( k_{\rm in},  q_{n})\,
 \frac{  e^{ i\, k_{\rm in} \left( x-L\right)}} 
{\sqrt{ \mathcal{V} (k_{\rm in}, q_n)}}
+ \,
\tilde t_n \, \Psi_- ( i \, \kappa ,  q_n) \,   e^{- \,\kappa \left( x-L\right) } \,,\nonumber \\
%%%%%%%%%%%%%%%%%%%%%%%%%%%%%%
%%%%%%%%%%%%%%%%%%%%%%%%
& k_{{\rm in}}  = \sqrt{ 2\,m\,E - q_n^2}\,, \quad
%%%%%%%%%%%%%%%%%%
\mathcal{V} (k_{\rm in},  q_n)  \equiv   
 |\partial_{k_{\rm in}} \varepsilon_{2d} (k_{\rm in}, q_n)|
=  \frac{ k_{\rm in} } {m}\,, \quad
%%%%%%%
\quad \kappa = \sqrt{ 2\,m\,E + q_n^2}\,,
\nn  
& k_{\rm mid}  = \sqrt{ 2\,m\,|E-V_0|  - q_n^2} \,,
\quad \kappa_{\rm mid} = \sqrt{ 2\,m\,|E-V_0|  + q_n^2}\,.
\end{align}
%%%%%%%%%%%%%%%%%%%%%%%%%
Here, the magnitude of the group-velocity, captured by $ \mathcal{V} (k_{\rm in},  q_n)$, is needed to define the scattering matrix comprising the transmission and reflection amplitudes.
It is important to note that, for $V_0 > E$, the Fermi level within the potential barrier lies within the valence band and we must use the valence-band wavefunctions in that region --- this input is captured by the usage of the Heaviside-theta functions.

The boundary conditions can be obtained by integrating the equation, $\mathcal{H}_{2d}^{tot}\, \Psi^{\rm{tot}} = E \, \Psi^{\rm{tot}} $, over a small interval in the $x$-direction, around each of the points at $x=0$ and $x=L$. The procedure is performed twice. The results are obtained as the two components of the wavefunction and their $x$-derivatives being continuous at the boundaries, as shown below:
%%%%%%%%%%%%%%%%%%%
\begin{align}
\label{eqbdy1}
& \phi_L (0) = \phi_M (0) \text{ and } 
%%%%%%%%%%%%%%
\partial_x \phi_L  (x) \big \vert_{x=0}
= \partial_x \phi_M  (x) \big \vert_{x=0} \,;\nn
%%%%%%%%%
%%%%%%%%%%%%%%%%%%%%%
& \phi_M (L) = \phi_R (L) \text{ and } 
%%%%%%%%%%%%%%
\partial_x \phi_M  (x) \big \vert_{x= L}
= \partial_x \phi_L  (x) \big \vert_{x= L} \,.
\end{align}
%%%%%%%%
These conditions are sufficient to guarantee the continuity of the flux of the probability-current density along the $x$-direction. From the matching of the wavefunction and its derivatives, we have $2\times 2= 4$ matrix-equations from the two boundaries. For 2d QPCBs, each of these matrix-equations can be separated into two components, since each wavevector has two components. Therefore, we have $ 4\times 2 = 8 $ equations for the eight undetermined coefficients,
$ \lbrace r_n , \, \tilde r_n,\, \alpha_n, \, \beta_n , \, \tilde \alpha_n ,\,
\tilde \beta_n ,  \, t_n, \tilde t_n  \rbrace $. The explicit analytical expressions for $t$ and $r$ are extremely long and, hence, we refrain from showing them here.

We would like to alert the reader that the evanescent-wave solutions do not contribute to the reflection and transmission coefficients, as they decay off as we move away from the barrier-junctions. Hence, the transmission and reflection coefficients at an energy $E$ are given by 
\begin{align}
T( E ,  V_0,\phi) = | t_n( E, V_0 )|^2 \text{ and }
R ( E ,  V_0,\phi) = | r_n( E, V_0 )|^2 ,
\end{align}
respectively, where $\phi = \tan^{-1} \left(  q_n/ k_{\rm in} \right)$
is the incident angle of the incoming plane wave.
Figs.~\ref{fig2dless} and \ref{fig2dmore} illustrate the polar plots of $ T (E, V_0,\phi)$ and $ R (E, V_0,\phi) $, as functions of $\phi$, for some representative values of $V_0$, $L$, and $E$. Instead, we represent their characteristics via Figs.~\ref{fig2dless} and \ref{fig2dmore}.
For the special case of $E \rightarrow 0 $, the explicit expressions take the form of
%%%%%%%%%%%%%%%%%%%%%%%%
\begin{align}
\label{eqe0}
T( E ,  V_0,\phi) = 
\frac{4 \, E \sec ^2 \phi 
\, \csc ^2 \big(L\, \sqrt{ 2\, m \,V_0} \big )} {V_0}
+ \order{E^2} \text{ and }
%%%%%%%%%%%%%%%%%
R ( E ,  V_0,\phi) =
1-\frac{4 \,E \sec^2 \phi \,\csc ^2
\big ( L \,\sqrt{ 2\, m \,V_0} \big )}
{V_0} + \order{E^2}\,.
\end{align} 
In the same limit, we get the transmission coefficient to be $ {\cos^2 \phi} / \left [
{1-\sin^2 \phi \cos ^2 \big (L  \,V_0 \big )} \right ]$
for (monolayer) graphene \cite{geim}, taking the Hamiltonian to be $ k_x \, \sigma_x + k_y \,\sigma_y $.

Let us compare the results obtained for a 2d QBCP with those for a 2d electron-gas. For normal metals, we have only one electronic band to consider, which the Fermi energy will intersect (irrespective of the height of the barrier). Using the continuity of the wavefunctions and their $x$-derivatives at the two ends of the barrier, we can easily find the transmission amplitude to be given by
\begin{align}
t_n (E, V_0)  = \begin{cases}
\frac{2 \, i \,  k_{\rm mid}\, k_{\rm in}}
{\left( k_{\rm mid}^2 + \, k_{\rm in}^2\right) \sin \left( k_{\rm mid}\, L \right )
+2  \, i \, k_{\rm mid}\, \, k_{\rm in} \cos \left ( k_{\rm mid} \, L \right)}
%%%%%%%%%%
& \text{ for } E \leq V_0 \\
%%%%%%%%%%%%%%%%%%%%%%%%%%%%%%%%%%%%%
\frac{  -\, 2 \, \kappa_{\rm mid}\, k_{\rm in}}
{ i \,
\left(  k_{\rm in}^2 - \kappa_{\rm mid}^2 \right) 
\sinh \left( \kappa_{\rm mid}\, L \right )
- 2  \,  \kappa_{\rm mid}\, \, k_{\rm in} 
\cosh \left ( \kappa_{\rm mid} \, L \right)}
%%%%%%%%%%
& \text{ for } E > V_0
\end{cases} \,.
\label{eqtval}
\end{align}
Figs.~\ref{fig2delecless} and \ref{fig2delecmore} illustrate the polar plots of the transmission and reflection coefficients for an electron-gas to provide a comparison with the case of 2d QBCPs.

%%%%%%%%%%%%%%%%%%%%%%%%%%%%%%%%%%%%%%%%%%%%%%%
\subsection{Transmission coefficients, conductivity, and Fano factors}

%%%%%%%%%%%%%%%%%%%%%%%%%%%%%%%%%%%%%%%%%%%%%%%%%%%
\begin{figure*}[t]
\subfigure[]{\includegraphics[width = 0.9 \textwidth]{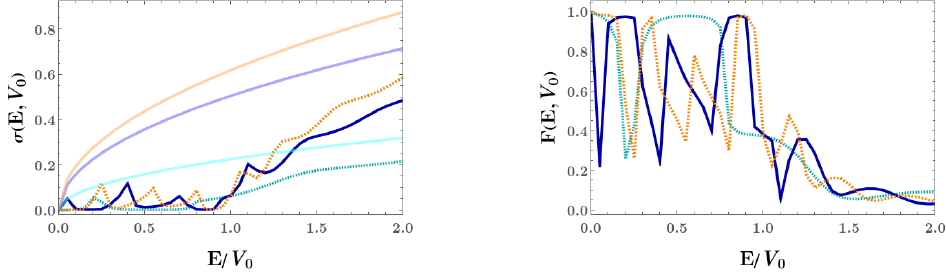}}
\subfigure[]{\includegraphics[width = 0.9 \textwidth]{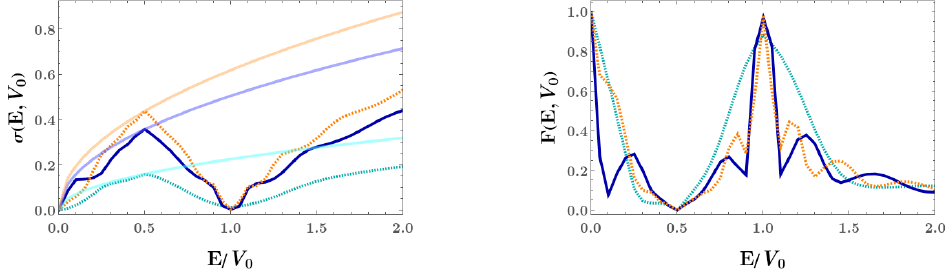}}
\subfigure{\includegraphics[width = 0.6 \textwidth]{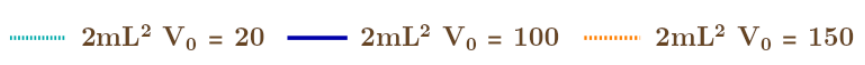}}
\caption{Plots of the conductivity ($\sigma$) and Fano factor ($F$), as functions of $E/V_0$, for various values of $2\, m \, L^2 \,V_0$. The subfigures (a) and (b) correspond to 2d QBCP and 2d electron-gas, respectively. Here, $\sigma$ and $F$ are unitless. \textcolor{black}{The light-coloured curves in the conductivity plots represent the Sharvin conductivity, as defined below Eq.~\eqref{eqsh}. In particular, the light-cyan, light-blue, and light-orange curves represent the cases when $2\,m\,L^2\,V_0 $
take the values of $20$, $100$, and $150$, respectively.} 
\label{figfano2d}
}
\end{figure*}
%%%%%%%%%%%%%%%%%%%%%%%

Let us assume $ W $ to be very large such that $q_n$ can effectively be treated as a continuous variable. Using $ k_{\rm in} = \sqrt{ 2 \,m \,E } \, \cos \phi $,
$ k_y = \sqrt{ 2 \,m \,E } \, \sin \phi$, and
$ dk_y = \sqrt{ 2\, m \, E }\; |\cos \phi| \, d\phi $ (using the continuum limit when $W \gg L$), we consider
the zero-temperature limit and  an applied  voltage of very small magnitude. Remembering that $ \hbar = 1 $ in the natural units, the conductance is given by [cf. Eq.~\eqref{eqland}]
\begin{align}
G (E,V_0) & = \frac{e^2} {2\,\pi} \sum_{n} |t_n|^2
= 
\frac{e^2} {2\,\pi\, \Delta  q} \sum_{n} |t_n|^2 \,(q_n-q_{n-1})
\quad \left (\text{where } \Delta  q = \frac{ 2\,\pi} {W}  \right )\nn
%%%%%%%%%%%%%%%
& \rightarrow \frac{e^2\, W} { 4 \,\pi^2} 
\int_{- \,\sqrt{2\, m\, E}}^{\sqrt{2\, m\, E}} dk_y\, T(E, V_0, k_y)
=
\frac{e^2\, W\,  \sqrt{ 2 \,m \,E } } { 4\,\pi^2} 
\int_{- \pi/2}^{\pi/2} d\phi \,
\cos\phi \, T(E, V_0, \phi)\,.
\end{align}
We note that the conductivity is given by \cite{beenakker}
\begin{align}
\sigma (E,V_0)  & =  \frac{ L} {W} \, G  (E,V_0)
=
\frac{e^2\, L\,  \sqrt{ 2 \,m \,E } } { 4\,\pi^2} 
\int_{- \pi/2}^{\pi/2} d\phi \,\cos\phi \, T(E, V_0, \phi)\,.
\end{align}
%%%%%%%%%%%%%%%%%
In Fig.~\ref{figfano2d}(a), we illustrate the conductivity and
the Fano factors [cf. Eq.~\eqref{eqfano}], as functions of $E/V_0$ , for some values of $2\, m\, L^2 \, V_0$. Side by side, the corresponding curves [generated using Eq.~\eqref{eqtval}] for a 2d electron-gas are provided in Fig.~\ref{figfano2d}(b), for the sake of comparison. From Eq.~\eqref{eqe0}, we note that $ \sigma \rightarrow 0 $ and $F \rightarrow 1$ in the limit $E \rightarrow 0 $. We can identify this behaviour from the plots. We note that, for graphene, we get the conductivity and the Fano factor to be
\begin{align}
& \frac{e^2 \, L \, E} {2\, \pi^2 }\,
\sec \left(L \, V_0\right) 
\left [ 2 \sec \big (L \, V_0\big )
+\tan ^2\big (L \,  V_0 \big) 
\ln \left(\tan ^2\left(\frac{L \,V_0}{2}\right)\right)\right ]
   \text{ and } \nn &
%%%%%%%%%
\frac{\tan^2\left(L \, V_0\right) \sec \left(L \, V_0\right) }
{4} \;
\frac{
\left \lbrace \cos \big (2\, L\, V_0 \big )-5 \right \rbrace
\sec \left(L \, V_0\right) \;
\ln \left(\tan^2 \big (\frac{L \, V_0}{2}\big )\right)
   -12 }
   {  2 \sec \left(L \, V_0\right)
+\tan^2\big (L \, V_0\big) 
\ln \left(\tan^2\left(\frac{L \,V_0}{2}\right)\right)}\, ,
\end{align}
respectively, for the limit $E \rightarrow 0 $.

%Let us also study the $ V_0 \rightarrow \infty$ limit, in which the transmission coefficient approximates to
%\begin{align}
%t=
%\frac{2 i k_{\text{in}} \left(\kappa ^2 k_{\text{in}}^2 \csch
%\left(L \sqrt{V_0}\right)
%-\kappa ^2 \, k_y^2 \csc \left(L
%   \sqrt{V_0}\right)+ \kappa ^2 \, k_y^2 \csch \left(L \sqrt{V_0}\right)
%   + k_y^4 \csc \left(L\,\sqrt{V_0}\right)\right)}
%   {\sqrt{V_0} \left(k_y^2+i \kappa  k_{\text{in}}\right)^2}\,.
%\end{align}
%This means that the transmission probability, $T = |t|^2$, contains the oscillatory components of $\csc \left(L\,\sqrt{V_0}\right)$
%as well as the decaying components of $\csch \left(L\,\sqrt{V_0}\right)$.

The fundamental upper bound of the ballistic conductance is given by the Sharvin conductance ($G_{\rm{sh}} $), which is equal to the conductance quantum $e^2 / (2\,\pi)$ (in natural units), multiplied by the number of quantum channels able to carry the electrons through the contact \cite{sharvin, brandbyge}. In other words, we have
\begin{align}
\label{eqsh}
& G_{\rm{sh}} (E,V_0)  =
 \frac{e^2\, W} {2\,\pi} \,\frac{1} {2\,\pi}
\int_{-\infty}^{\infty} dk_y\, T_{\rm ideal}(E, V_0, k_y),
%%%%%%%%%%
 \text{ where }
 T_{\rm ideal}(E, V_0, k_y)  = \Theta( k_F - |k_y|) 
\text{ and } k_F = \sqrt{ 2\, m\, E}\
%%%%%%%%%%%%%%%%%%%
\nn \Rightarrow & \;
G_{\rm{sh}} (E,V_0)  = \frac{e^2\, W\,  \sqrt{ 2\, m\, E}} { 2\,\pi^2} \,.
\end{align}
A comparison of $\sigma (E,V_0)$ with $\sigma_{\rm{sh}} (E,V_0) \equiv   
\frac{ L} {W} \, G_{\rm{sh}}  (E,V_0) $ is also provided in Fig.~\ref{figfano2d}.

%%%%%%%%%%%%%%%%%%%%%%%%%%%%%%%%%%%%%%%%%%%%%%%%%%%%%%%%%%%%%%
\section{3d QBCP}
\label{sec3dmodel}

A QBCP in a 3d system harbours a fourfold band-crossing, where the four bands form a four-dimensional representation of the underlying lattice-symmetry group \cite{MoonXuKimBalents}. The resulting $\left(\mathbf{k} \cdot \mathbf{p} \right)$ Hamiltonian, for the particle-hole symmetric system, can be written by using a representation of the five $4\times 4$ Euclidean Dirac matrices, $\lbrace \Gamma_a \rbrace $, as \cite{murakami, Herbut}
 \begin{equation}
 \mathcal{H}_{3d}^{kin}(k_x,k_y,k_z) = \frac{1} {2\,m}
 \sum_{a=1}^5 d_a(\mathbf {k}) \,  \,\Gamma_a   \,.
\label{bare}
 \end{equation}
%%%%%%%%%%%%%%%%
The $\Gamma_a$-matrices form one of the (two possible) irreducible four-dimensional Hermitian representations of the five-component Clifford algebra, defined by the anticommutator $\{ \,\Gamma_a, \,\Gamma_b \} = 2\, \delta_{ab}$.
The five anticommuting gamma-matrices can always be chosen such that three are real and two are imaginary \cite{murakami,igor12}. In the representation used here, $(\Gamma_1, \Gamma_3, \Gamma_5)$ are real and $(\Gamma_2, \Gamma_4 ) $ are imaginary:
\begin{align}
\Gamma_1 = \sigma_1 \otimes \sigma_0 \,, \quad  \Gamma_2 = \sigma_2 \otimes \sigma_0 \,,
\quad \Gamma_3 = \sigma_3 \otimes \sigma_1 \,, \quad \Gamma_4 = \sigma_3 \otimes \sigma_2 \,,\quad
\Gamma_5 = \sigma_3 \otimes \sigma_3 \,.
\end{align}
The five functions, $ \lbrace d_a(\mathbf{k}) \rbrace $, are the real $ \ell =2$ spherical harmonics, with the following structure:
\begin{align}
& d_1(\mathbf k)= \sqrt{3}\,k_y\, k_z \, , \,\,  
d_2(\mathbf k)= \sqrt{3}\,k_x \,k_z \,, \,\, 
d_3(\mathbf k)= \sqrt{3}\,k_x\, k_y  , \quad
 d_4(\mathbf k) = \frac{\sqrt{3}}{2}(k_x^2 -k_y^2) \,, \,\, 
d_5(\mathbf k) =
\frac{1}{2}\left (2\, k_z^2 - k_x^2 -k_y^2 \right ) .
\end{align}
%%%%%%%%%%%%%%%%%%%%%
The energy eigenvalues are given by $\pm \, \varepsilon_{3d}(k_x,k_y,k_z)$, where
\begin{align}
\varepsilon_{3d}(k_x,k_y,k_z) =
\frac{  k_x^2 + k_y^2 +k_z^2 } {2\,m} \,.
\end{align}
The ``$+$" and ``$-$" signs, as usual, refer to the conduction and valence bands, respectively. Each of these bands is doubly degenerate.

A set of orthonormal eigenvectors is given by the following:
%%%%%%%%%%%%%%%%%%%%%%%%%
\begin{align}
\Psi_{-,1}^T  &= \frac{1} { \mathcal{N}_{-,1} } 
\Bigl(
-\frac{(k_x+ i\, k_y) \left(k+k_z\right)}
{(k_x- i\, k_y)^2}
\qquad
\frac{ i\, \left( k+3 \,k_z\right)}
{\sqrt{3} \,(k_x- i\, k_y)}
\qquad
-\frac{ i\, \left(-2\, k_z \left(k+k_z\right)+k_x^2+k_y^2\right)}
{\sqrt{3}\, (k_x- i\, k_y)^2}
\qquad
1 \Bigr), \nonumber\\
 \Psi_{-,2}^T
&=  \frac{1} { \mathcal{N}_{-,2} } 
\Biggl( 
\frac{(k_x+ i\, k_y) \left(k-k_z\right)} 
{(k_x- i\, k_y)^2}
\qquad
-\frac{ i\, \left(k-3\, k_z\right)}
{\sqrt{3} \,(k_x- i\, k_y)}
\qquad
-\frac{ i\, \left(2\, k_z \left( k-k_z\right)+k_x^2+k_y^2 \right)}
{\sqrt{3}\, (k_x- i\, k_y)^2}
\qquad 1 \Biggr) ,\nn 
 \Psi_{+,1}^T &= 
\frac{1} { \mathcal{N}_{+,1} } 
\Biggl(
-\frac{ i\, \left(k+k_z\right)}  {\sqrt{3}\, (k_x- i\, k_y)}
\quad
\frac{ k-k_z} {k_x+ i\, k_y}
\qquad 
1
\qquad
-\frac{ i\, \left(2\, k_z \left(k_z-k \right)+
k_x^2+k_y^2 \right)}
{\sqrt{3}\, (k_x+ i\, k_y)^2} \Biggr) ,\nn
 \Psi_{+,2}^T &= 
\frac{1} { \mathcal{N}_{+,2} } 
\Biggl(
 \frac{ i\, \left(k-k_z\right)}{\sqrt{3}\, (k_x- i\, k_y)}
\qquad 
-\frac{k+k_z}{k_x+ i\, k_y}
\qquad
1
\qquad
-\frac{ i\, \left(2 \,k_z 
\left(k+k_z\right)+k_x^2+k_y^2\right)}
{\sqrt{3}\, (k_x+ i\, k_y)^2}
\Biggr),
\end{align}
%%%%%%%%%%%%%%%%%%%%%%
where $k= \sqrt{k_x^2  + k_y^2 +k_z^2}$. The ``$+$" (``$-$") subscript indicates an eigenvector corresponding to the positive (negative) eigenvalue. The symbols $\frac{1} { \mathcal{N}_{\pm,1}}$ and $\frac{1} { \mathcal{N}_{\pm,2}}$ denote the corresponding normalization factors.

The 3d system is modulated by a piecewise-constant electric-potential barrier of
strength $V_0 $ and width $L$. Here, we choose the $z$-axis as the transport direction, and place the chemical potential at an energy $E >0$ in the regions outside the potential barrier. Since the system is isotropic, it does not matter which axis we choose. But, in the end, it is just a matter of convenience while transforming to the conventional spherical polar coordinates. Therefore, we have
\begin{align}
V ( z ) = \begin{cases} 
V_0 &  \text{ for } 0 \leq z \leq L \\
0 & \text{ otherwise} 
\end{cases} .
\label{eqpot2}
\end{align}
As before, we need to consider the total Hamiltonian in the position space as
\begin{align}
\mathcal{H}_{3d}^{tot} &=
 \mathcal{H}_{3d}^{kin}(- i\,\partial_x, - i\,\partial_y, - i\,\partial_z)+V(z) \,.
\end{align} 
All this amounts to replacing $k_x$ by $k_z$  in Fig.~\ref{figbands}.

%%%%%%%%%%%%%%%%%%%%%%%%%%%%%%%%%%%%%%%%
\subsection{Formalism}

We consider the tunneling in a slab of height and width $W$.
Again, we assume that the material has a sufficiently large width $W$ along each of the two transverse directions, such that the boundary conditions are irrelevant for the bulk response. Here, we impose periodic boundary conditions on a position-space wavefunction, $ \tilde \Psi^{{\rm tot}}(x,y,z)$, translating into
$ \tilde \Psi^{\rm{tot}}(x,0,z) =
 \tilde \Psi^{\rm{tot}}(x, W,z)$ and $\tilde \Psi^{\rm{tot}}( 0,y,z) = 
\tilde \Psi^{\rm{tot}}(W , y,z) $.
%%%%%%%%%%%%%%%%%%%%%%%%%%%%%%%%
The  transverse momentum, $ \mathbf k_\perp=(k_x , k_y)$, is conserved, and its components are quantized. Imposing periodic boundary conditions, we demand that
$
k_x =\frac{ 2\,\pi\,n_x } {W} \equiv q_{n_x}$ and $k_y =\frac {2\,\pi\,n_y} {W} \equiv q_{n_y}$,
where $ \mathbf n \equiv \lbrace n_x, n_y \rbrace  \in \mathbb{Z}$. Consequently, the full wavefunction can be written as
\begin{align}
\tilde  \Psi^{\rm{tot}} (x,y,z, q_{n_x}, q_{n_y}) =
\text{const.}
\times \tilde  \Psi_{ \mathbf{n}}(x)\, 
 e^{  i \, \left( q_{n_x} x \, + \,  q_{n_y}  y \right ) }\,.
\end{align}
%%%%%%%%%%%%%%%%%%%%%%%%%%%%%%%%%%%%%%%
For any mode with a given value of the transverse-momentum magnitude, defined as $ k_\perp = \sqrt{q_{n_x} ^2  +  q_{n_y} ^2}$ (with $ k_\perp \leq \sqrt{2\, m\,E} $), we can determine the $z$-component of the wavevectors of the incoming, reflected, and transmitted waves (denoted by $k_{{\rm in}}$), by solving $
4 \, m^2 \, E^2 = \left( k_z^2 + k_\perp^2 \right)^2 \Rightarrow k_z^2
 = \pm \, 2 \, m\, E - k_\perp^2 \,.$
%%%%%%%%%%%%%%%%%%%%%%%
In fact, in the regions $z<0$ and $z>L$, this relation leads to the following four solutions: 
\begin{align}
\label{eqksol2}
k_z = \pm \,\sqrt{2 \, m\, E - k_\perp^2} \text{ and }
k_z = \pm \, i\, \sqrt{2 \, m\, E + k_\perp^2} \,.
\end{align}
%%%%%%%%%%%%%%%%%%%%%%%
As we have understood by now from the 2d case, this implies the presence of evanescent waves, corresponding to the imaginary solutions for $k_z$.

We will follow the same procedure as described for the 2d QBCP. Without any loss of generality, we consider the transport of one of the degenerate positive-energy states, $\Psi_{+,1}$, corresponding to electron-like particles for one of the two degenerate conduction bands. Here, the Fermi level outside the potential barrier is adjusted to a value $E =\varepsilon_{3d} (k_x,k_y, k_z) $.
In this case, a scattering state $\sim \tilde  \Psi_{\mathbf n} (z) \, e^{i \, ( q_{n_x} x \, + \, q_{n_y} y )}$, labelled by the vectorial index $\mathbf n$, is
constructed in a piecewise fashion as
\begin{align}
 \tilde \Psi_{ \mathbf n} (z)=
 \begin{cases} 
 \tilde \phi_L (z)  & \text{ for }  z <0   \\
\tilde  \phi_M  (z) & \text{ for } 0 \leq z \leq L  \\
\tilde \phi_R (z)  &  \text{ for } z > L 
\end{cases} ,
\end{align}
%%%%%%%%%%%%%%%%%%%%%%%%%%%%%%%%%%%%%%%%%%%%5
where
%%%%%%%%%%%%%%%%%%%%%%%%%%%%%%%%%%%%%%%%
\begin{align}
 & \tilde \phi_L (z) = \frac{   
 \Psi_{+,1} ( q_{n_x},q_{n_y},  k_{\rm in}) 
 \, e^{ i\, k_{\rm in} \, z }
+  
\sum \limits_{s =1,2}
 r_{\mathbf n,s} \,\Psi_{+,s} ( q_{n_x},q_{n_y}, - \,k_{{\rm in}} ) 
 \, e^{- i\, k_{\rm in} \, z }
}
{\sqrt{ \tilde{ \mathcal{V} }(k_{\rm in}, \mathbf n) }}
%%%%%%%%%%%%%%%
+ 
\sum \limits_{s =1,2}
{\tilde r}_{\mathbf n,s} \,
\Psi_{-,s} (  q_{n_x},q_{n_y}, -\, i \, \kappa ) 
 \, e^{ -\, \kappa  \, |z| }
\, ,\nonumber \\
%%%%%%%%%%%%%%%%%%%%%%%%%%%%%%%%%%%%%%%%%%%%5
%%%%%%%% mid region %%%%%%%%%%%%%%%%%%
& \tilde \phi_M  (z) = \sum \limits_{s =1,2}
\Bigg[  
 \alpha_{\mathbf n,s} \,\Psi_{+,s} ( q_{n_x},q_{n_y}, k_{\rm mid}) \,
 e^{ i\,k_{\rm mid}\,  z } 
 + 
 \beta_{\mathbf n,s} \,\Psi_{+,s} ( q_{n_x},q_{n_y}, -\,k_{\rm mid}) \,
 e^{- i\,k_{\rm mid}  \, z } 
\nn & \hspace{ 2.5 cm } + \,
 %%%%%%%%%%
\tilde \alpha_{\mathbf n,s} \,\Psi_{-,s} ( q_{n_x},q_{n_y}, i\, \kappa_{\rm mid}) \,
 e^{ - \,\kappa_{\rm mid}\,  z } 
 + 
\tilde  \beta_{\mathbf n,s} \,\Psi_{-,s} ( q_{n_x},q_{n_y}, -\,i\,\kappa _{\rm mid}) \,
 e^{ \kappa_{\rm mid}  \, z }  
\Bigg]   \Theta\left( E-V_0  \right)  \nonumber\\
& \hspace{ 1.25 cm}
%%%%%%
 + \sum \limits_{s =1,2}
 \Bigg[
 \alpha_{\mathbf n,s} \,\Psi_{-,s}  ( q_{n_x},q_{n_y}, k_{\rm mid})\,
 e^{ i\,k_{\rm mid}\,  z } 
 + 
 \beta_{\mathbf n,s} \,\Psi_{-,s}  ( q_{n_x},q_{n_y}, -\,k_{\rm mid}) \,
 e^{- i\,k_{\rm mid}  \, z }
\nn & \hspace{ 2.75 cm } + \,
 %%%%%%%%%%
\tilde \alpha_{\mathbf n,s} \,\Psi_{+,s} ( q_{n_x},q_{n_y}, i\, \kappa_{\rm mid}) \,
 e^{ - \,\kappa_{\rm mid}\,  z } 
 + 
\tilde  \beta_{\mathbf n,s} \,\Psi_{+,s} ( q_{n_x},q_{n_y}, -\,i\,\kappa _{\rm mid}) \,
 e^{ \kappa_{\rm mid}  \, z }   
   \Bigg]  \,\Theta\left( V_0-E  \right) \,,\nonumber \\
 %%%%%%%%%%%%%%%%%%%%%%%%%%%%%%%%%%%%%%%%%%%%%%%%%%%%%%%%%%%
%%%%%%%%%%%%%%%%%%%%%%%%%%%% 
& \tilde \phi_R (z) = 
\frac{ \sum \limits_{s =1,2}
 t_{\mathbf n,s} \,\Psi_{+,s} ( q_{n_x},q_{n_y}, k_{\rm in} ) 
 } 
{\sqrt{ \tilde{ \mathcal{V} }  (k_{\rm in}, \mathbf n)}}
\, e^{ i\, k_{\rm in} \left( z-L\right)}
+ 
\sum \limits_{s =1,2}
{\tilde t}_{\mathbf n,s} \,
\Psi_{-,s} (  q_{n_x}, q_{n_y}, i \, \kappa ) 
 \, e^{   \kappa  \,(z-L)} ,\nonumber \\
%%%%%%%%%%%%%%%%%%%%%%%%
%%%%%%%%%%%%%%%%%%%%%%%%%%%
& k_{{\rm in}}  = \sqrt{ 2\,m\,E - k_\perp^2}\,, \quad
\tilde{\mathcal{V}} (k_{\rm in},  \mathbf n)  \equiv   
 |\partial_{k_{\rm in}} \varepsilon_{3d} 
 ( q_{n_x}, q_{n_y}, k_{\rm in} )|
=  \frac{ k_{\rm in} } {m}\,, \quad
%%%%%%%
\quad \kappa = \sqrt{ 2\,m\,E + k_\perp^2}\,,
\nn  
& k_{\rm mid}  = \sqrt{ 2\,m\,|E-V_0|  - k_\perp^2} \,,
\quad \kappa_{\rm mid} = \sqrt{ 2\,m\,|E-V_0|  + k_\perp^2}\,.
\end{align}
Here, the magnitude of the group velocity, defined by $ \tilde{ \mathcal{V} }(k_{\rm in}, \mathbf n) $, is needed to define the reflection and transmission amplitudes appearing in the unitary scattering matrix. 

%%%%%%%%%%%%%%%%%%%%%%%%%%%%%%%%%%%%%%%%%%%%%%%%%%
\begin{figure}[t]
\includegraphics[width = 0.75 \textwidth]{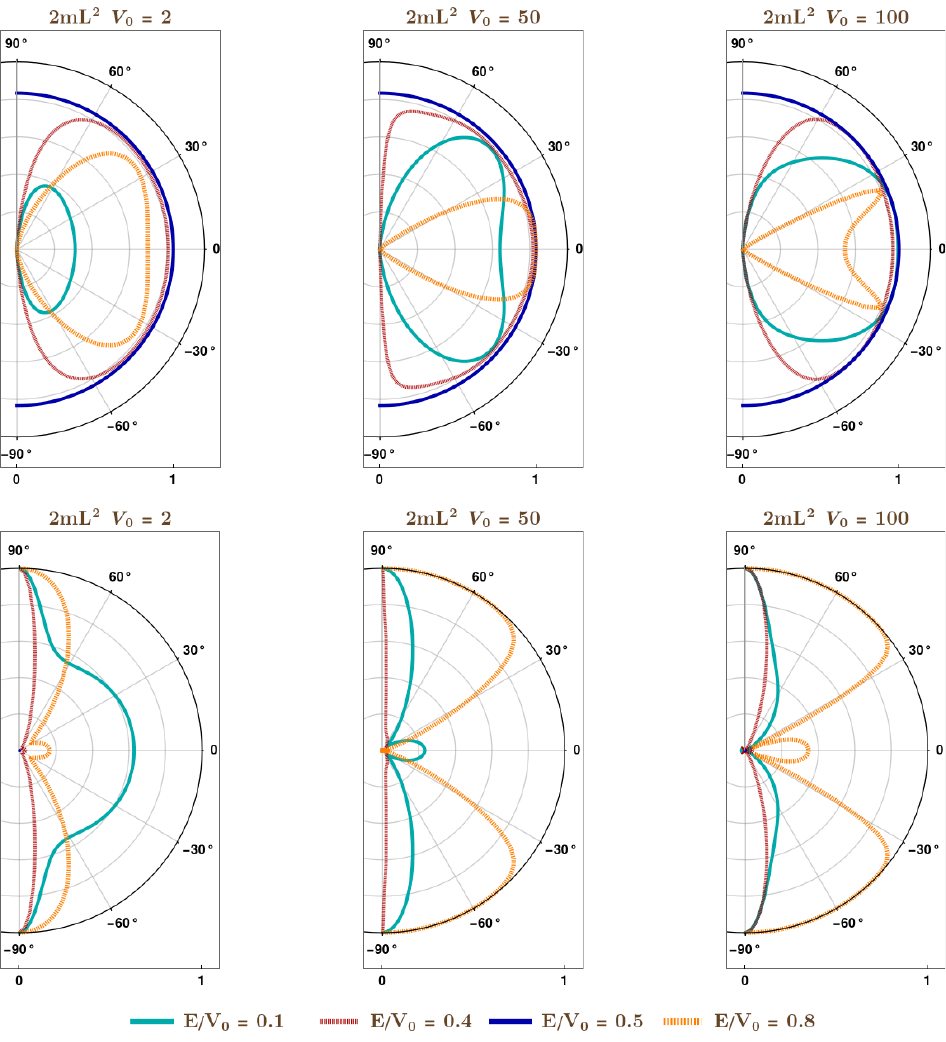}
\caption{3d QBCP: The polar plots show the transmission coefficient, $ T (E, V_0,\theta) \big \vert_{E < V_0}$ (upper panel), and the reflection coefficient $ R (E, V_0,\theta) \big \vert_{E \leq V_0}$ (lower panel), as functions of the incident angle $\theta$, for some representative values of $2\, m\, L^2\,V_0$. The values of $E$ are shown in the plot-legends.
\label{fig3dless}}
\end{figure}
%%%%%%%%%%%%%%%%%%%%%%%%

%%%%%%%%%%%%%%%%%%%%%%%%%%%%%%%%%%%%%%%%%%%%%%%%%%
\begin{figure}[t]
\includegraphics[width = 0.75 \textwidth]{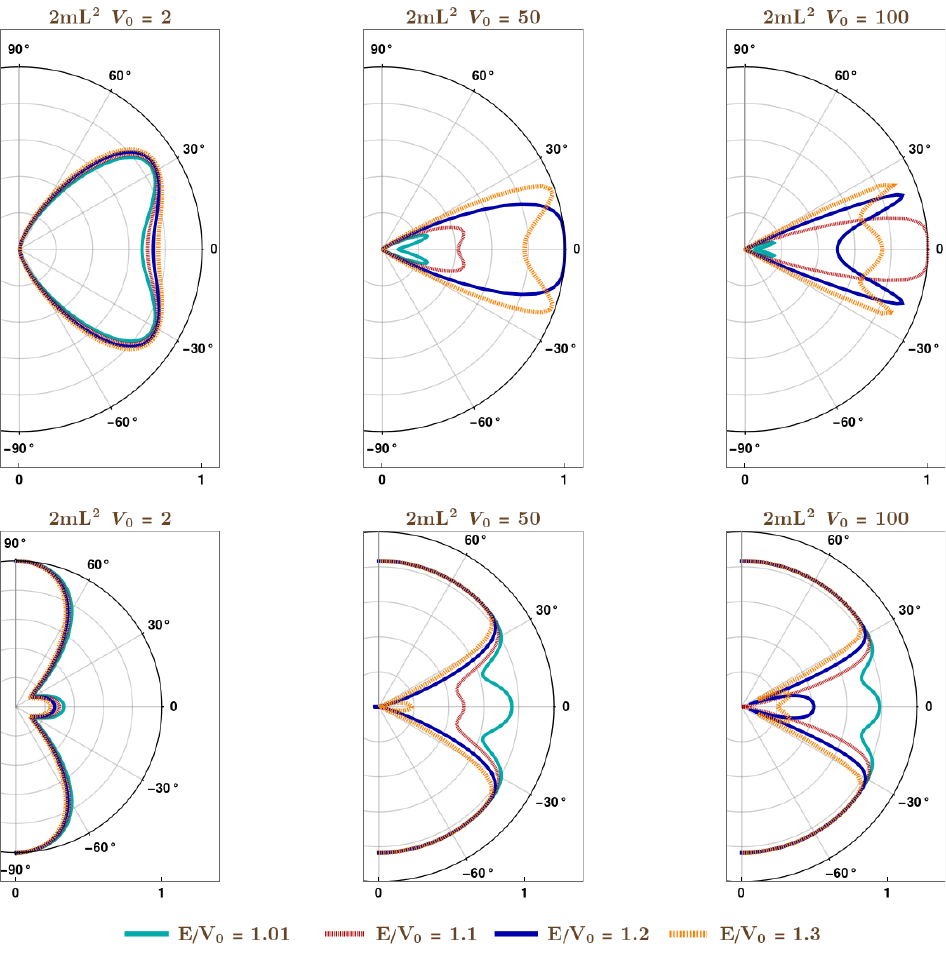}
\caption{3d QBCP: The polar plots show the transmission coefficient, $ T(E, V_0,\theta) \big \vert_{E > V_0}$ (upper panel), and the reflection coefficient $ T (E, V_0, \theta) \big \vert_{E > V_0}$ (lower panel), as functions of the incident angle $\theta$, for some representative values of $2\, m\, L^2\,V_0$. The values of $E$ are shown in the plot-legends.
\label{fig3dmore}}
\end{figure}
%%%%%%%%%%%%%%%%%%%%%%%%

%%%%%%%%%%%%%%%%%%%%%%%%%%%%%%%%%%%%%%%%%%%%%%%%%%%
\begin{figure}[t]
\subfigure[]{\includegraphics[width = 0.9 \textwidth]{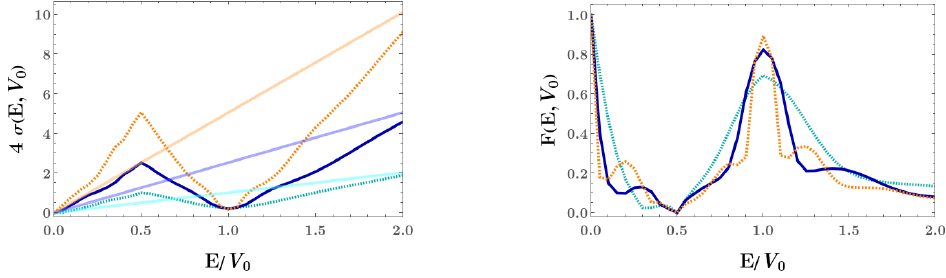}}
\subfigure[]{\includegraphics[width = 0.9 \textwidth]{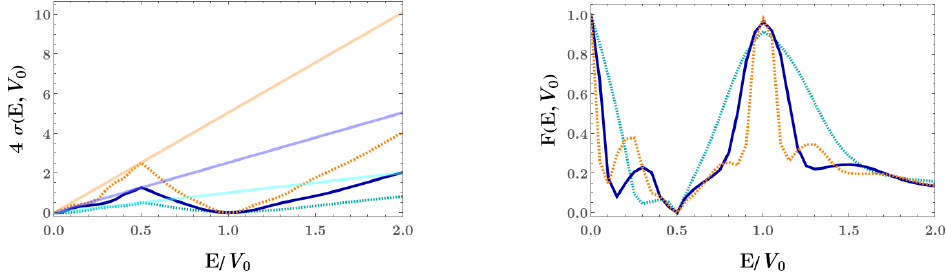}}
\subfigure{\includegraphics[width = 0.6 \textwidth]{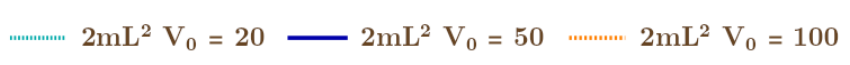}}
\caption{Plots of the conductivity ($\sigma$) and Fano factor ($F$), as functions of $E/V_0$, for various values of $2\, m \, L^2 \,V_0$. The subfigures (a) and (b) correspond to 3d QBCP and 3d electron-gas, respectively. While $\sigma$ has the units of eV, $F$ is unitless. \textcolor{black}{The light-coloured curves in the conductivity plots represent the Sharvin conductivity, as defined below Eq.~\eqref{eqsh2}. The light-cyan, light-blue, and light-orange curves represent the cases when $2\,m\,L^2\,V_0 $ take the values of $20$, $100$, and $150$, respectively.}
\label{figfano3d}}
\end{figure}
%%%%%%%%%%%%%%%%%%%%%%%

The boundary conditions at $z=0$ and $ z=L$ lead to
%%%%%%%%%%%%%%%%%%%
\begin{align}
\label{eqbdy2}
&   \tilde \phi_L (0) = \tilde \phi_M (0) \text{ and } 
%%%%%%%%%%%%%%
\partial_z \tilde  \phi_L  (z) \big \vert_{z=0}
= \partial_z \tilde \phi_M  (x) \big \vert_{z=0} \,;\nn
%%%%%%%%%
%%%%%%%%%%%%%%%%%%%%%
& \tilde \phi_M (L) = \tilde \phi_R (L) \text{ and } 
%%%%%%%%%%%%%%
\partial_z \tilde  \phi_M  (z) \big \vert_{ z= L}
= \partial_z \tilde \phi_L  (z) \big \vert_{ z= L} \,.
\end{align}
%%%%%%%%
These conditions correspond to the continuity of the flux of the probability-current density along the $z$-direction. From the matching of the wavefunction and its derivatives, we have $2\times 2 = 4$ matrix-equations from the two boundaries.
For 3d QPCBs, each of these matrix-equations can be separated into four components, since each wavevector comprises four components. Therefore, we have $ 4 \times 4 = 16 $ equations for the sixteen undetermined coefficients, captured by
$ \lbrace r_{\mathbf{n}, 1} , \,
 \tilde r_{\mathbf{n}, 1},\, \alpha_{\mathbf{n}, 1}, \, \beta_{\mathbf{n}, 1}, \, 
 \tilde \alpha_{\mathbf{n}, 1} ,\,
\tilde \beta_{\mathbf{n}, 1} ,  \, t_{\mathbf{n}, 1}, \tilde t_{\mathbf{n}, 1}  \rbrace 
\, \cup \,
\lbrace r_{\mathbf{n}, 2} , \, \tilde r_{\mathbf{n}, 2},\, \alpha_{\mathbf{n}, 2}, \, \beta_{\mathbf{n}, 2}, \, \tilde \alpha_{\mathbf{n}, 2} ,\,
\tilde \beta_{\mathbf{n}, 2} ,  \, t_{\mathbf{n}, 2}, \tilde t_{\mathbf{n}, 2}  \rbrace 
$. Again, since the explicit analytical expressions for $t_{\mathbf{n}, 1}$, $t_{\mathbf{n}, 2}$, $ r_{\mathbf{n}, 1}$, and $ r_{\mathbf{n}, 2 }$ are extremely long, we refrain from showing them here. Instead, we represent their characteristics via Figs.~\ref{fig3dless} and \ref{fig3dmore}.

The reflection and transmission coefficients, at an energy $E$, are given by 
\begin{align}
R( E ,  V_0,\theta ) = | r_{\mathbf n, 1}( E, V_0)|^2 
+ | r_{\mathbf n, 2}( E, V_0 )|^2
 \text{ and }
T( E ,  V_0,\theta ) = | t_{\mathbf n, 1}( E, V_0 )|^2  
+ | t_{\mathbf n, 2}( E, V_0 )|^2 \,,
\end{align}
respectively,
where $\theta = \tan^{-1} \left(  k_\perp / k_{\rm in} \right)$
defines the incident angle of the incoming wave. Although we have not provided the explicit expressions
for $t_{\mathbf{n}, 1}$ and $t_{\mathbf{n}, 2}$, one can check that $| t_{\mathbf n, 2}( E, V_0 )| = 0$ when the incident state is assumed to be $ \Psi_{+,1} $.\footnote{Similarly, $| t_{\mathbf n, 1}( E, V_0 )| = 0$ when the incident state is assumed to be $ \Psi_{+,2} $.}
Figs.~\ref{fig3dless} and \ref{fig3dmore} illustrate the polar plots of $ T (E, V_0,\theta)$ and $ R (E, V_0,\theta) $, as functions of $\theta $, for some representative values of $V_0$, $L$, and $E$. For the sake of presentation, we have artificially extended the $\theta$-values to negative angles in the polar plots. If one wants to compare the results obtained for a 3d QBCP with those for a 3d electron-gas, Eq.~\eqref{eqtval} and Figs.~\ref{fig2delecless}--\ref{fig2delecmore} are to be considered (i.e., the same as the 2d electron-gas).

%%%%%%%%%%%%%%%%%%%%%%%%%%%%%%%%%%%%%%%%%%%%%%%
\subsection{Transmission coefficients, conductivity, and Fano factors}

Again, we assume $ W $ to be very large, such that $q_{n_x}$ and $ q_{n_y}$ can effectively be treated as continuous variables. 
Using $ k_{\rm in} = \sqrt{ 2 \,m\,E }  \cos \theta$,
$ k_x =  \sqrt{ 2\, m \,E }  \sin \theta \cos \phi $, and
$ k_y =  \sqrt{ 2\, m \,E }  \sin \theta \sin \phi  $, we get
$ dk_x \, dk_y =
d\phi\,d\theta \, m\, E  \,|\sin (2\, \theta)| $. Hence,
in the zero-temperature limit, the conductance is given by [cf. Eq. \eqref{eqland}]
%%%%%%%%%%%%%%%%%%%%%%%%
\begin{align}
G (E,V_0) & = 2\times \frac{e^2} {2\,\pi} \sum_{n} |t_{\mathbf n, 1}|^2
%%%%%%%%%%%%
\rightarrow \frac{e^2\, W^2} { 4\,\pi^3 } 
\int_{0}^{\sqrt{2\, m\, E}} dk_\perp \, k_\perp
\int_0^{2\pi} d\phi\; T(E, V_0, k_\perp) \nn
%%%%%%%%%%%%%%%%%%%
& = \frac{e^2} {\pi} \,
\frac{ W^2 \,(2\, m\, E)} { 2 \,\pi }
\int_{0}^{\pi/2} d\theta \, |\sin \theta \, \cos \theta |\,  T(E, V_0, \theta)
=
\frac{ e^2\, W^2 \, m\, E} { 2\, \pi^2}
\int_{0}^{\pi/2} d\theta \, 
|\sin (2\theta ) |\;  T(E, V_0, \theta)\,.
\end{align}
In order to account for the twofold degeneracy (since we have two independent conduction bands), we have included an extra factor of two.
We note that the conductivity is given by
\begin{align}
\sigma (E,V_0)  & =  \frac{ L} {W^2} \, G  (E,V_0)
= \frac{ e^2\, L \, m\, E} { 2\,\pi^2}
\int_{0}^{\pi/2} d\theta \, 
|\sin (2\theta ) |\;  T(E, V_0, \theta)\,.
\end{align}
%%%%%%%%%%%%%%%%%%
In Fig.~\ref{figfano3d}(a), we illustrate the conductivity and
the Fano factors [using Eq.~\eqref{eqfano}], as functions of $E/V_0$ , for some values of $2\, m\, L^2 \, V_0$. Side by side, the corresponding curves [generated using Eq.~\eqref{eqtval}] for a 3d electron-gas are provided in Fig.~\ref{figfano3d}(b) for the sake of comparison.
\textcolor{black}
{In this case, the Sharvin conductance is given by
\begin{align}
\label{eqsh2}
& G_{\rm{sh}} (E,V_0)  = \frac{e^2\, W^2 \, m\, E} { 4\,\pi^2} \,.
\end{align}
A comparison of $\sigma (E,V_0)$ with $\sigma_{\rm{sh}} (E,V_0) \equiv   
\frac{ L} {W^2}  \, G_{\rm{sh}}  (E,V_0) $ is also provided in Fig.~\ref{figfano3d}.
}

%%%%%%%%%%%%%%%%
\section{Summary and outlook}
\label{secsum}

In this paper, we have revisited the problem of the tunneling of quasiparticles for 2d and 3d QBCP semimetals, mainly to correct the implementations of the boundary conditions by including parts featuring evanescent waves. The inclusion of these evanescent waves, arising from a QBCP's quadratic-in-momentum dispersion, were missed in our earlier work \cite{ips_tunnel_qbcp}. With the correct number of wavevector-solutions in place, we have proceeded to compare the transport-characteristics (obtained using the Landauer-Büttiker formalism) with those of normal electron-gases. The results also enable us to contrast with those for other nodal-point semimetals exemplified by graphene \cite{geim, allain}, pseudospin-$1$ (or triple-point) semimetals \cite{fang, zhu, ips3by2}, Rarita-Schwinger-Weyl semimetals \cite{ips3by2}, and Weyl/multi-Weyl semimetals \cite{mansoor, deng2020, ips-aritra, ips-jns}. The conclusion that QBCPs do not exhibit Klein tunneling remains valid (where Klein tunneling is characterized by $T = 1$ at normal incidence), unlike the Dirac quasiparticles of graphene, Weyl/multi-Weyl semimetals, and triple-point fermions. Such differences can be observed by designing appropriate experimental set-ups involving the QBCPs. The extra evanescent waves are the artifacts of a dispersion whose momentum dependence along the propagation direction is of an order higher than the linear-order cases. Semi-Dirac semimetals \cite{banerjee, ips-abs-semid}, bilayer graphene \cite{geim}, and multi-Weyl semimetals \cite{deng2020, ips-aritra, ips-jns} provide examples of such nonlinear-in-momentum dispersive bands.

In the future, it will be worthwhile to determine the nature of Josephson junctions constructed out of the 2d and 3d QBCPs, and determine the nature of the Andreev bound states with sub-gap energies. This exercise has been carried out for graphene, Weyl/multi-Weyl semimetals, semi-Dirac nodal points, pseudospin-1 semimetals, and Rarita-Schwinger-Weyl semimetals \cite{krish-moitri, debabrata-krish, ips-abs-semid, ips_jj_rsw, ips-jj-spin1}. The computations will be significantly harder than the nodal-point semimetals (considered till now in the literature), because the presence of the extra evanescent waves (resulting from the quadratic-in-momentum dispersion) will lead to solving polynomial equations resulting form the determinant of $8$-dimensional and $16$-dimensional matrices for the 2d and 3d QBCPs, respectively, for a Josephson junction modelled as a delta-function potential \cite{ips-abs-semid, ips-qpcp-delta}.

\bibliography{biblio}

\end{document}